\documentclass[acmsmall]{acmart}

\setcopyright{none}

\usepackage[utf8]{inputenc}
\usepackage{outlines}
\usepackage{listings}
\usepackage{xspace}
\usepackage{semantic}
\usepackage{cleveref}
\usepackage{mathtools}
\usepackage{longtable}
\usepackage{booktabs}

\lstdefinelanguage{JavaScript}{
  morekeywords={typeof, new, true, false, catch, function, return, null, catch, switch, var, if, in, while, do, else, case, break, async, await, require, const, let},
  morecomment=[s]{/*}{*/},
  morecomment=[l]//,
  morestring=[b]",
  morestring=[b]'
}

\definecolor{editorOcher}{rgb}{1, 0.5, 0} 
\definecolor{editorGreen}{rgb}{0, 0.5, 0} 

\lstdefinestyle{js} {%
  basicstyle={\footnotesize\ttfamily},
  frame=b,
  xleftmargin={0.75cm},
  numbers=left,
  stepnumber=1,
  firstnumber=1,
  numberfirstline=true,	
  identifierstyle=\color{black},
  keywordstyle=\color{blue}\bfseries,
  ndkeywordstyle=\color{editorGreen}\bfseries,
  stringstyle=\color{editorOcher}\ttfamily,
  commentstyle=\color{brown}\ttfamily,
  language=JavaScript,
  alsodigit={.:;},	
  tabsize=2,
  showtabs=false,
  showspaces=false,
  showstringspaces=false,
  extendedchars=true,
  breaklines=true,
  literate=%
  {Ö}{{\"O}}1
  {Ä}{{\"A}}1
  {Ü}{{\"U}}1
  {ß}{{\ss}}1
  {ü}{{\"u}}1
  {ä}{{\"a}}1
  {ö}{{\"o}}1
}

\newcommand{\eat}[1]{}
\newcommand{\ie}[0]{\textit{i.e.}\xspace}
\newcommand{\eg}[0]{\textit{e.g.}\xspace}

\theoremstyle{definition}

\newcommand{\rawLineRef}[1]{Line~\ref{line:#1}}
\newcommand{\lineRef}[1]{(\rawLineRef{#1})}

\begin{document}

\title[This is not the End]{This is not the End: Rethinking Serverless Function Termination} 

\author{Kalev Alpernas}
\orcid{0000-0001-6979-8880}
\affiliation{
  \institution{Tel Aviv University}
  \country{Israel}
}
\email{kalevalp@tauex.tau.ac.il}

\author{Aurojit Panda}
\affiliation{
  \institution{NYU}
  \country{USA}
}

\author{Mooly Sagiv}
\affiliation{
  \institution{Tel Aviv University}
  \country{Israel}
}

\begin{abstract} 
Elastic scaling is one of the central benefits provided by serverless platforms, and
requires that they scale resource up and down in response to changing workloads.
Serverless platforms scale-down resources by terminating previously launched
instances (which are containers or processes). The serverless programming 
model ensures that terminating instances is safe \emph{assuming} all application
code running on the instance has either completed or timed out. Safety thus depends
on the serverless platform's correctly determining that application processing is complete.

In this paper, we start with the observation that current serverless platforms do not
account for pending asynchronous I/O operations when determining whether application
processing is complete. These platforms are thus unsafe when executing programs that
use asynchronous I/O, and incorrectly deciding that application processing has terminated
can result in data inconsistency when these platforms are used.  We show that 
the reason for this problem is that current serverless semantics couple termination 
and response generation in serverless applications. We address this 
problem by proposing an extension to current semantics that decouples response 
generation and termination, and demonstrate the efficacy and benefits of our proposal
by extending OpenWhisk, an open source serverless platform. 
\end{abstract}

\maketitle

\section{Introduction}
Serverless computing (also referred to as \textbf{F}unctions-\textbf{a}s-\textbf{a}-\textbf{s}ervice FaaS) is a recently proposed execution model that has been widely adopted by cloud providers, and operators of large compute cluster. The serverless execution model is designed to ensure that \emph{correct applications} can be safely auto-scaled, \ie to ensure that in the absence of application bugs the runtime can \emph{add} or \emph{remove} resources from the application without impacting application semantics. The execution model provides safe auto-scaling by requiring that applications be structured as collections of \emph{function} which are only executed in response to external events such as web requests, or file system write; by limiting the semantics of individual functions so that any state accessible across requests must be explicitly stored in a remote cloud-provider service (\eg a database); and by requiring that functions process any external events within a finite time-bound. Given these constraints, it is easy to see that additional resources can be safely added for an application whenever a new event occurs, and assigned resources can be safely removed once an event has been processed. The requirement that all events are processed within a finite time-bound ensures that application bugs cannot result in resource leaks. The ability to easily build and deploy elastically-scalable applications, which only use as many resources as required to serve the current workload, has been a significant driver for the wider adoption of serverless computing.

Serverless runtimes, including ones implemented by cloud providers such as AWS~\cite{aws-lambda} and open source projects such as OpenWhisk~\cite{openwhisk-github-repo}, rely on these application semantics for allocating and freeing up application resources: these runtimes assume that application resources can be safely allocated whenever a \emph{new event occurs}, and safely freed whenever \emph{event processing is completed}. However, to implement this policy serverless runtimes need to determine when a function has completed processing an event. Making such a determination is challenging when using languages such as JavaScript and Python that support \emph{asynchronous operations} (\eg asynchronous I/O) which run concurrently with the function and can complete \emph{after the function returns}. In existing serverless runtimes, events are modeled as REST calls (\ie as web requests), and a function is considered to have completed processing an event either when it produces a response for this REST call, or when event processing times out. As we show in \S\ref{sec:curr-semantics}, when using asynchronous operations, application developers need to be careful to ensure that all asynchronous operations have completed before responding to a REST call, and if they are not then applications can exhibit erroneous behaviour. Furthermore, the need to wait for all asynchronous operations to complete also adds to the latency of the REST call, which in many applications translates to a worse user experience. 

A majority of serverless applications are written in languages such as Python and JavaScript that support asynchronous I/O operations~\cite{loring2017semantics}, and make use of libraries that use asynchronous I/O to interact with cloud services~\cite{aws-sdk}. Most of these libraries were originally designed for use in programs running on traditional servers, rather than for serverless applications. As we discuss later in \S\ref{sec:curr-semantics}, the widespread use of asynchronous I/O operations in libraries used by a majority of serverless applications has been a frequent source of bugs when writing serverless application. In this paper we take the position that the mechanism currently used by serverless runtimes in order to determine when event processing is completer needlessly complicates serverless applications, and reduces the utility of serverless applications. 

In \S\ref{sec:prop-semantics} we propose an alternate set of semantics where we decouple when a function responds to an event, and when event processing is completed. We argue that these new semantics simplify the task of writing new serverless applications, and improve the performance of serverless applications. Then in \S\ref{sec:implementation}, we demonstrate the practical viability of our semantics by modifying OpenWhisk to implement these semantics, and porting serverless applications to run on top of OpenWhisk.

\section{Background}\label{sec:background}

\subsection{Serverless Computing}
Serverless computing (sometimes referred to as functions-as-a-service) is an emerging
execution model for cloud computing applications. Serverless applications are comprised
of a set of self-contained functions, and a set of event triggers which specify what function
should be executed in response to a real world event. For example, a serverless web application
might be comprised of a function $W$ that is triggered by web requests, and another $S$ that is triggered by
changes to storage. When a web-request arrives the cloud platform executes (invokes) function $W$ in an
isolated container. The serverless execution model requires developers to assume that each
function invocation processes exactly one event, that each invocation runs for a bounded 
amount of time after which it is terminated, and that local state is not shared between
function invocations. Functions can however access cloud services, including databases and filesystems,
and serverless applications use these services to share state between functions.

As has been discussed in prior work~\cite{jangda2019formal}, serverless platforms introduce
new execution semantics. These new semantics simplify resource management, scaling, and reduce the
cost of deploying applications on the cloud, and have thus been a major impetus for the adoption
of serverless computation. However, these new execution semantics pose significant challenges
when writing serverless applications. This is because most applications rely on external libraries, \eg to
interact with cloud services and perform operations such as serialization. However, most of these libraries
are not designed for serverless environments, and instead target application running on traditional
servers. As a result using these libraries in serverless environments can violate safety requirements
for these libraries, and result in application bugs.

\subsection{Asynchronous Programming in JavaScript}
In the rest of this paper we focus on Serverless applications written in JavaScript. These applications
form a significant majority~\cite{serverlessCommunitySurvey} of all deployed Serverless applications. However, the issues we outline
also impact applications written in other languages and out approach can thus be generalized to Serverless
applications written in other languages.

JavaScript serverless applications are executed using the Node.js~\cite{dahl2012node} runtime, and must use asynchronous,
non-blocking I/O when accessing temporary local storage or communicating over the network (\eg to use other cloud
services)~\cite{tilkov2010node}. Applications can rely on either callbacks or \emph{promises}~\cite{liskov1988promises} to wait for the completion of asynchronous
events. 

Promises are objects that represent an ongoing asynchronous computation. A
promise may be in one of three states---\emph{pending}, \emph{fulfilled}, or
\emph{rejected}. Fulfilled promises represent successful completion of the
asynchronous computation; rejected promises represent a failed asynchronous
computation; pending promises are promises that are neither fulfilled nor
rejected, i.e., they have not yet completed. \emph{Resolved} promises are
promises that are no longer pending, that is, they have completed execution
either by fulfillment or rejection. Resolved promises hold a resolved value,
which is the result of the asynchronous operation. In the case of fulfilled
resolved promises they the value is the result of the successful asynchronous
computation; in the case of rejected promises the value is an error object,
representing the error that occurred in the execution.

Promise objects are a useful mechanism for working with asynchronous
operation, especially compared with the traditional mechanism available in
JavaScript---callbacks, and the resulting aptly-named callback-hell~\cite{gallaba2015don}. Promises
can be chained using \texttt{then}, \texttt{catch}, and \texttt{finally}
methods. The chain methods are used to register event handlers, function that
run when the promise is resolved. The \texttt{then} methods accepts two
parameters: a \texttt{resolved} handler and \texttt{fulfilled} handler, which
run when the promise is resolved or fulfilled, respectively. The
\texttt{catch} method accepts a handler which runs when the promise is
resolved with rejection, and the \texttt{finally} method accepts a handler
which will run when the promise is resolved, regardless of whether the promise
is fulfilled or rejected. 

All three chain methods return a promise, allowing further promises to be
chained and so on. Calling a chain method on a resolved promise will cause the
handler to run according to the chaining rules---calling \texttt{then} or
\texttt{finally} on a fulfilled promise will cause the handler to run
immediately, whereas calling \texttt{catch} on a fulfilled promise will not run the
handle, but calling \texttt{catch} on a rejected promise will cause the
handler to run immediately, etc.

To create more complex control flows, promises can be combined using several
\emph{promise combinator} functions. The combinators accept a list of promises
and return another promise whose resolution depends on the state of the
constituent promise. These combinators include, \texttt{Promise.all()}, which
returns a promise that fulfills when all input promises fulfill and
rejects when any of them rejects, \texttt{Promise.allSettled()} which fulfills
when all input promises resolve, \texttt{Promise.any()}, which fulfills when 
any of the constituent promises fulfills or rejects when all of them reject 
(the opposite of \texttt{Promise.all()}), and \texttt{Promise.race()} which 
fulfills when the first input fulfills and rejects when the first input 
rejects.

\paragraph{Promise graphs}

In this work we rely on the formalism of \emph{promise graphs}, introduced by
\citet{madsen2017model} to describe asynchronous executions in JavaScript programs. Promise graphs describe the causal relation between
different promises in a JavaScript program, providing a convenient visual
representation of the control flow of a JavaScript program and the
asynchronous operations started by the program. Promise graphs can be used to
detect and analyze a variety of problems, such as race conditions and
starvation (\cite{alimadadi2018finding}).

Nodes in the promise graph represent promises created by the program, and
edges in the graph represent causal relation, e.g., fulfillment of promises
and execution of a \texttt{then} handler. In this work we use a simplified version of promise graphs to
illustrate the control flow of code example, and demonstrate race conditions
as well as issues resulting from those race conditions and our proposed
solutions for these issues.

\section{Problem Statement}\label{sec:problem-statement}

\subsection{The termination problem}\label{sec:termination-problem}
Serverless functions have a request-response pattern. A requestor, typically a remote user, sends a request to the system. The request event is sent to a serverless function. The serverless function processes the event,
and replies to the requestor with a response. The response contains the
function output, and potentially additional information, such as execution
metadata (run time, memory usage, etc.).

Modern language runtimes consist of a myriad of components, many of which are
running in separate threads in parallel to the main execution. These may
include, for example, garbage collector threads~\cite{steele1975multiprocessing}. 
One of the guarantees of serverless systems is that they do not bill users for idle time, i.e., time in which no events are processed. Additionally, serverless system need to be able to downscale resources when those resources are no longer needed. For both these reasons, the function needs to 
determine the moment event processing is finished. Since other threads may be present, 
waiting for true idle is not applicable. 

Instead, serverless platforms chose the moment a response to the event request
is produced as the marker for the end of event processing. However, in the 
presence of asynchronous I/O, or more broadly, general multithreading, stopping the running container the 
moment a response is produced may cause some parts of the program not to run
or run inconsistently. The resulting system may exhibit anything from hard to debug inconsistent behaviour to data corruption.
We call this problem \emph{the termination problem of serverless function execution}.

\subsection{A Running Example}\label{sec:example}

In this section, we present an example program used to illustrate the ideas
presented in this paper.

\begin{figure}
\footnotesize
\begin{lstlisting}[style=js]
const cloudProvider = require('...') (*@ \label{line:cloud-provider-require} @*)
const db = cloudProvider.DB()        (*@ \label{line:cloud-provider-db} @*)

let val, hash                        (*@ \label{line:global-vars} @*)

function computeHash(val) {...}      (*@ \label{line:hash-func-impl} @*)

function main (event) => {     (*@ \label{line:main-serverless-function} @*)
  val = event.val                    (*@ \label{line:extract-val} @*)
  hash = computeHash(val)            (*@ \label{line:compute-hash} @*)

  db.connect(...) // returns a promise         (*@ \label{line:connect} @*)
    .then((con) => con.write({val, hash})      (*@ \label{line:store} @*)
    .catch((err) => ...)                       (*@ \label{line:catch-err} @*)

  return db.connect(...)                       (*@ \label{line:db-second-connect} @*)
         .then((con) => con.read(...))         (*@ \label{line:db-read} @*)
         .then((stored) => ({stored, hash}))   (*@ \label{line:return-stored-n-hash} @*)
  
}
\end{lstlisting}
\Description{Code for the running example}
\caption{A serverless function that computes some hash of the input. The
function stores the original and the hash in a database. The function reads a
value from a database, and returns the stored value and the hash to the user.
Both database operations are simultaneous asynchronous I/O
operations.}\label{code:main-handler}
\end{figure}

\begin{figure}
    \centering
    \includegraphics[width=0.95\textwidth]{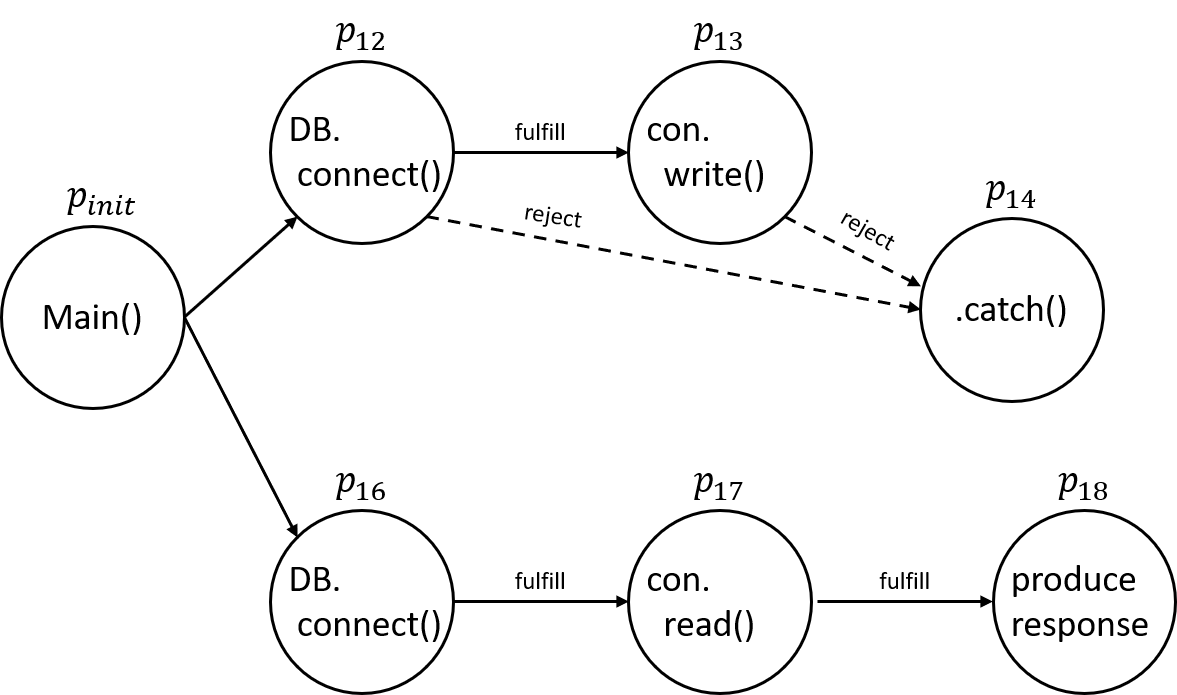}
    \Description{Promise graph for the running example}
    \caption{The promise graph of the main function. Note that there is no
    path in the graph from the db-write promise to the function return. The
    two promise chains create a race condition.}
    \label{fig:example-promise-graph}
\end{figure}

\Cref{code:main-handler} shows a program implementing a simple serverless
function that computes and stores user-provided inputs in a cloud database 
service, and then reads a second value from the database and returns it to
the user. To ensure value integrity, the application computes a hash and
stores it with the value when writing to the database, and also includes the
hash in user responses.

The function defines two global variables \texttt{hash}, and \texttt{val}. When
invoked the function first reads the input value \lineRef{extract-val},
and computes its hash \lineRef{compute-hash}. 

Next, the function begins an asynchronous operation to write to the database 
(\rawLineRef{connect}--\rawLineRef{catch-err}). This begins by first asynchronously creating
a database connection \lineRef{connect}; and then asynchronously writing to the database \lineRef{store}.
The function also registers an error handler to handle any errors that occur when writing to the
database \lineRef{catch-err}. The entire write operation (starting with \texttt{db.connect(...)} \lineRef{connect})
occurs asynchronously, and the main function continues to execute concurrently.

Concurrently with the database write, the main function also begins an asynchronous
operation to read from the database
(\rawLineRef{db-second-connect}--\rawLineRef{return-stored-n-hash}). This proceeds similarly,
except the promise that will eventually resolve to the value read from the database is 
returned as the result of the function \lineRef{return-stored-n-hash}. The serverless platform
waits for this promise to be resolved and then sends the resulting value back to the user.

\Cref{fig:example-promise-graph} shows the promise graph
(\cite{madsen2017model}) of the main serverless function. Nodes in the promise graph represent promise objects created in the program, and edges represent causal relation. The labels above the nodes denote the line in the code in which the promise is produced (e.g., $p_{12}$ is a promise created in line 12 of the program), and the labels inside nodes represent the operation that the promise performs (e.g., \texttt{con.write()} is the asynchronous I/O operation of writing to the database). Labels above edges describe which resolution caused this transition, i.e., was the promise fulfilled or rejected. We omit from the promise graph intermediate promises created by \texttt{.then()} calls.

There is a node in
the graph for the start of the execution ($p_{init}$), followed by a fork,
with a path for each of the asynchronous tasks in the program. The first path
consists of a database connection ($p_{12}$), the database write operation ($p_{13}$), and the error handling
\texttt{catch()} call ($p_{14}$). The second path consists of another database
connection ($p_{16}$), the database read
operation ($p_{17}$) and finally the production of the response that will be sent back to the user ($p_{18}$).

The promise graph describes immediate causality between asynchronous
operations created during the execution of the system. In the graph in
\Cref{fig:example-promise-graph} there is no path from the database write
operation and the function return (or vice-versa). The function returns an unresolved promise; when the promise resolves, the resolved value will be sent back to the user as a response. Because there is no causal path between the database write and the return we can conclude that in
this function there is no guarantee that the database write operation will
terminate before the response value is produced. In other words, we have a race
condition between the database write operation and the response being produced.

The serverless function terminates when the response is produced, i.e., when
promise $p_{18}$ is resolved. Consequently, we have a race condition between 
the write operation and serverless function termination. This raises
the question: \emph{Will the database be updated?} As we show in
\cref{sec:curr-semantics}, the answer to this question is that the behaviour is undefined.

\subsection{The real-world impact of the termination problem}
We know from other domains that undefined semantics, similar to the ones caused
by the serverless termination problem, can make it challenging to produce correct
code. However, does this extend to serverless, \ie does the
termination problem actually impact real world serverless developers? To address
this question we looked at two months of Stack Overflow questions about JavaScript 
applications deployed on AWS Lambda and found 7 different questions which seem
to stem from the termination problem
\cite{stackof-q1,stackof-q2,stackof-q3,stackof-q4,stackof-q5,stackof-q6,stackof-q7}.

In all of these questions asynchronous operations do not seem to complete as expected,
and the questions cover a range of services that operate asynchronously (S3, AWS IoT, Nodemailer,
Axios), and the initial questions often try to understand \emph{why the
library does not behave correctly}. 

The core challenge these questions reveal is the following: modern applications, especially
those that use cloud services, depend on a large number of external libraries. At present
most of these libraries are designed to work with applications running on normal servers,
and tested on these normal servers. Additional care is thus necessary when using these 
libraries in serverless applications specifically to ensure that a function does not respond
\emph{before} all asynchronous operations have completed. This is however impossible in general
(libraries may not return a promise that captures all outstanding asynchronous operation) and
suboptimal when possible since it increases response latencies. Additionally, as the forum posts
we cited show it adds to program complexity, thus motivating an alternate approach where we change
serverless execution semantics to eliminate the termination problem.

In the next section we formalize serverless semantics to better explain the 
cause of the termination problem, and then in \S\ref{sec:prop-semantics} we
propose a modified set of semantics that provides a more intuitive execution and 
termination model for serverless functions \emph{without} impacting the performance
or scalability of serverless functions.

\section{The \emph{Current} Semantics of Serverless Function Termination}\label{sec:curr-semantics}

In this section we provide a formal model describing serverless function
executions. We present two model variants---the \emph{single-execution model},
and the \emph{function-reuse model}---and provide formal semantics for both
variants. 
Finally, we use the models to demonstrate several important consequences of 
the termination problem.
In the next section (\S\ref{sec:prop-semantics}), we propose a modification to
the semantics that fixes these problems, and in \S\ref{sec:implementation} we
describe an implementation of the fixed semantics in a cloud platform.

\subsection{Formal semantics of the single-execution model}

We start with the \emph{single-execution model}, a simplified model of
serverless function execution. In this model a new serverless function is
created whenever a new request arrives. A function only processes a single
event, and is never run again. Local state at one serverless function is not
accessible by any other function.

In practice serverless platforms use a model that allows function reuse.
However, because each invocations of a serverless function may start in a new
environment, developers are expected to write applications that work
correctly in the single-execution model.

\begin{figure}
\begin{flalign*}
& \Sigma:(Var \rightarrow Val) \times (\mathcal{E}_{id}\cup\{f,d\}) \times 2^{\mathcal{PR}}) &\\
& \sigma = (M, E, Pr) &\\
& \sigma_{init} = (\emptyset, f, \emptyset) &\\
\end{flalign*}
\Description{Local state of a function}
\caption{The local state of a single serverless function.}\label{sem:state}
\end{figure}

Figure~\ref{sem:state} describes the local state of a single serverless
function. A state $\sigma$ is composed of three components: (i) $M$, a mapping
from memory location to values, (ii) $E$ the unique id of event currently being
processed, and (iii) $Pr$, a set of \emph{promises} representing
asynchronous operations running in the function.

The memory mapping is defined in the regular way.

The set $Pr\subseteq\mathcal{PR}$ of promises represents the currently
unresolved parallel/asynchronous tasks in the function with $\mathcal{PR}$
being the (infinite) set of all possible promises that can run in the system.
The set $Pr$ is initially empty.

The currently processed event $E$ represents the processing state of the serverless function. It can either store an event
identifier $e\in\mathcal{E}_{id}$, unique to each executing run, or
one of two special values, $f$ (as in free) representing a free serverless
function that can potentially process an event, and $d$ (as in done)
representing a terminal state of the serverless function.

The initial state of each function consists of an empty memory mapping, the
event state $f$ representing a `free' function, and an empty set of promises.

\begin{figure}
Start Event:
\[
<receive(v), (M, f, Pr)>\Longrightarrow(M[v/input], newEid(), Pr)
\]\vspace{5pt}

End Event:
\[
\inference{M(response)=v & e\in\mathcal{E}_{id}}{<respond(v), (M, e, Pr)>\Longrightarrow(M,d,Pr)}
\]\vspace{5pt}

Local Transition:
\[
\inference{e\in\mathcal{E}_{id}}{<C, (M, e, Pr)>\Longrightarrow(M', e, Pr)}
\]\vspace{5pt}

Start Asynchronous Task:
\[
<startAsync(p), (M, E, Pr)>\Longrightarrow(M, E, Pr\cup\{p\})
\]\vspace{5pt}

End Asynchronous Task:
\[
\inference{p\in Pr}{<resolve(p), (M, E, Pr)>\Longrightarrow(M, E, Pr\setminus\{p\})}
\]\vspace{5pt}
\Description{Semantics of the single-execution model}
\caption{Small-step operational semantics of the single-execution model of 
serverless function execution. $f$ and $d$ are event processing states
representing free and done respectively. The value $v$ sent as part of the End
Event is read from the reserved memory location $response$. In the Local
Transition rule, the memory state $M'$ is obtained by running the command $C$
with the memory state $M$ according to the semantics of the code running in
the serverless function. The detailed semantics of the language that the
serverless function is running are defined in the standard
way.}\label{sem:existing-simple}
\end{figure}

Figure~\ref{sem:existing-simple} describes the small-step operational
semantics rules for serverless functions running under the single-execution model.

The Start Event rule represents the start of a new serverless function
execution. The cause for this event is a start message sent to the function.
The event payload, $v$---the input to the serverless function---is stored in
the $input$ memory location. The function $newEid()$ produces a new unique
event id $e\in\mathcal{E}_{id}$. The id $e$ is then stored in the $E$
component of the post-state.

The premise for the rule requires that the event component of the state be
$f$.

The following lemma guarantees that in the single-execution model a single
function does not process multiple events simultaneously:

\begin{lemma}[Single event processing]\label{lem:single-event-processing}
A serverless function only processes a single event at any one time
\end{lemma}

The End Event rule represents the termination of the serverless function. The
end event sends a response value $v$ back to the originator of the event. The value $v$ is read from the memory location $response$. The
currently processed event component of the state is changed to $d$,
representing a terminal state of the execution.

\begin{lemma}[No function revival]\label{lem:no-function-revival}
Once a function has terminated it will never process an event again
\end{lemma}

The following theorem guarantees the single-execution property of the model:

\begin{theorem}[No function reuse]\label{thm:no-function-reuse}
Every serverless function processes at most one event
\end{theorem}

The theorem follows directly from lemmas~\ref{lem:single-event-processing} and
\ref{lem:no-function-revival}

The Start Asynchronous Task rule represents the start of a new asynchronous
task represented by the promise object $p$. The promise $p$ is added to the
set of promises $Pr$.

The End Asynchronous Task rule represents the resolution of promise $p$. We
require that the promise $p$ be a part of the set of promises $Pr$.

The Local Transition rule represents a step in the execution of the program. $M'$ is the memory state obtained from the memory state $M$ after performing the command $C$.
The details of
the execution are defined in the standard way. For a function to perform a
step it is necessary that the function process some event at the time of
execution, hence, the premise of the rule requires that $e$ be an execution id.

The following lemma guarantees that in the single-execution model the
serverless function does not perform any processing when idle:

\begin{lemma}[No idle run]
A function only performs computation when it is processing an event
\end{lemma}

We call any asynchronous code that has not finished running by the time the
function terminates according to the termination condition \emph{residual
execution}. 
We define as residual execution the set $Pr$ of promises in the post-state of the End rule (i.e., the promises available immediately after the application of the End rule). These represent asynchronous computation that has
not been concluded by the time event processing has ended.

The following lemma shows that residual executions may occur in serverless function under the single execution model:

\begin{lemma}[Residual executions exist]
The simple model of the current semantics admits function executions that
result in termination with a non empty set of promises as residual execution
\end{lemma}

We call the phenomenon of asynchronous operation not executing
\emph{broken promises}. The following theorem shows that broken promises
may occur in serverless functions running under the single execution model:

\begin{theorem}[Broken promises]
A promise $p$ which is part of the residual execution of a function will never
be resolved
\end{theorem}

\subsubsection{The impacts of broken promises}\label{sec:broken-promises-impacts}

The broken promises problem presents several challenges to developers writing
serverless applications. The main problem with broken promises is
\emph{correctness}, or lack thereof. If some parts of the application that the
developer expected to run do not run, the resulting program state is incorrect
w.r.t the developer's expectation. However, the correctness problem caused by
broken promises has several nuances that are important to observe. Broken
promises present a significant challenge to developers for two main reasons:
(i) their mere existence is a departure from the traditional semantics of
program execution---this occurrence is unique to serverless platforms, and
(ii) the behaviour of asynchronous tasks is timing dependent and prone to race
conditions, which results in potentially inconsistent recurrence of the
problem. We elaborate on broken promises and nondeterminism in \cref{sec:broken-nondet}.

\paragraph{Unexpected semantics}

Typically, language runtimes will wait for all asynchronous and parallel tasks
to finish before terminating the process. Developers have an expectation that
whatever code was started will run, barring any errors in the execution. A
model in which in an error-free execution asynchronous tasks are stopped at
some arbitrary point, and never complete their run is not natural to most
developers. When developing serverless applications developers reason about
and test serverless functions against the standard semantics.

\paragraph{Limited portability of traditional applications}

In traditional applications based on the request-response model, i.e.,
applications running on a server where the server is constantly available, it
is a common design pattern to send the response back to the user as soon as
the response is available, and finish in the background the rest of the
computation that was started by the request. The main reason for choosing this
design pattern is optimizing for user-perceived latency---delay in sending a
response to the user has impacts on the user experience.

This difference in programming models means that any attempt to port
traditional applications to a serverless setting will need to take into
account and mitigate the possibility of broken promises occurring in the
execution. Mitigating broken promises in the current model will necessitate
delaying the response to the user until all processing is done, which may
result in significant impact on the user experience. Under these
circumstances, we see the broken promises problem as a major hurdle holding
back serverless application adoption.

\subsection{Example run of the single-execution model}\label{sec:single-exec-model-example-run}

We demonstrate an execution of a request of the serverless function in the running example
(\S\ref{sec:example}).

\begin{center}
\footnotesize
\begin{longtable}{rlllll}
\caption{An execution of the serverless function in the running example under
the single-execution model. The example run consists of an invocation of the
serverless function with the input value 42. We omit from the execution 
intermediate promises created by the \texttt{then} calls. We denote the
computed hash value of $x$ by $\mathcal{H}(x)$ and the  stored value that the
program reads from the database by $\mathcal{S}$.}\label{table:simple-model-run}
\endfirsthead
\endhead
\toprule
\# & loc & command & state & unresolved promises & comment\\
\midrule
0 & l\ref{line:main-serverless-function} & main(event) &  & & \\
 &  &  & \texttt{val -> undefined} & & \\
 &  &  & \texttt{hash -> undefined} & & \\
 &  &  & \texttt{event -> \{val: 42\}} & & \\
\midrule
1 & l\ref{line:extract-val} & val = event.val &  & & \\
 &  &  & \texttt{val -> 42} & & \\
 &  &  & \texttt{hash -> undefined} & & \\
 &  &  & \texttt{event -> \{val: 42\}} & & \\
\midrule
2 & l\ref{line:compute-hash} & hash = computeHash(val) &  & & \\
 &  &  & \texttt{val -> 42} & & \\
 &  &  & \texttt{hash -> \(\mathcal{H}(42)\)} & & \\
 &  &  & \texttt{event -> \{val: 42\}} & & \\
\midrule
3 & l\ref{line:connect} & db.connect(\ldots{}) &  & \{$p_{12}$ (db.connect)\}\\
\midrule
4 & l\ref{line:db-second-connect} & db.connect(\ldots{}) &  & \{$p_{12}$ (db.connect), & connection\\
 &  &  &  & $p_{16}$ (db.connect)\} & started\\
\midrule
5 & l\ref{line:db-second-connect} & db.connect(\ldots{}) &  & \{$p_{12}$ (db.connect)\} & connection\\
 &  &  &  &  & established\\
\midrule
6 & l\ref{line:db-read} & conn.read(...) &  & \{$p_{12}$ (db.connect), & read \\
 &  &  &  & $p_{17}$ (con.read)\} & started \\
\midrule
7 & l\ref{line:db-read} & conn.read(...) &  & \{$p_{12}$ (db.connect)\} & read \\
 &  &  &  &  & finished \\
\midrule
8 & l\ref{line:return-stored-n-hash} & () => (\{stored, hash\}) &  & \{$p_{12}$ (db.connect)\} & response \\
&  &  &  & $p_{18}$ (produce response)\} & started \\
\midrule
9 & l\ref{line:return-stored-n-hash} & (\{$\mathcal{S}$, $\mathcal{H}(42)$\}) &  & \{$p_{12}$ (db.connect)\} & response \\
&  &  &  &  & produced \\
\bottomrule
\end{longtable}
\end{center}

Table~\ref{table:simple-model-run} describes the execution of a serverless
function call under the single-execution model. We denote the computed hash
value of $x$ by $\mathcal{H}(x)$ and the stored value that the program reads
from the database by $\mathcal{S}$.

The input value for the invocation is 42. The run starts from the initial
state of the function---the global variables \texttt{val} and \texttt{hash},
are initially \texttt{undefined}. The global variable \texttt{val} is updated
in step \#1, and the hash is computed in step \#2.

In \#3 the application starts a new asynchronous operation.
The \texttt{db.connect} call starts a connection with the database service.
The call returns a promise object that resolves to a connection object once
the connection has been established. The connection object can then be used to
perform database operations. The \texttt{db.connect} call sends a message to
the database service, and awaits a reply. Since this call is asynchronous
(i.e., non blocking), the mainline code of the function continues executing
while the reply is pending. Once the reply arrives, the promise is resolved
and the asynchronous code can continue executing.

However, as we will see next, this does not happen in the single-execution
model.

When the mainline code reaches \rawLineRef{db-second-connect} in step \#4, the
application starts a second asynchronous operation. In step \#4 a connection
is started, and in \#5 a connection is successfully created. Next, the \texttt{read} operation starts in step \#6. In step \#7 the \texttt{read} operation successfully
returns. In steps \#8-\#9 the response value---the object \texttt{(\{stored, hash\})}---is
evaluated to the value \{$\mathcal{S}$, $\mathcal{H}(42)$\}, where $\mathcal{S}$ is the 
value that returned from the \texttt{read} operation.

The value evaluated in step \#9 is the value resolved by the promise that is
returned from the function, and so this is the response sent back to the
caller. Recall that this is also the termination condition of the function.
The function is now considered done, and no further processing will
occur.

At the moment the main function returns, the response from the database server
has not yet arrived. As a consequence, any operation that were slated to occur
once the database connection was established had not yet had a chance to run.
Since the environment does not perform any further processing,
this means that the \texttt{con.write} operation (l\ref{line:store}) will
never be performed.

\subsubsection{Broken promises and nondeterminism}\label{sec:broken-nondet}

Whenever parallel execution or asynchronous I/O are involved, there is a risk
of race conditions making reasoning and debugging harder, and this case is no
different. The race condition between the read and write operations may result
in some cases where the write successfully concludes, and others where it does
not.

We demonstrate an alternative execution of the same request as in
\cref{table:simple-model-run} in which the write is successful.

\begin{center}
\footnotesize
\begin{longtable}{rlllll}
\caption{An alternative execution of the running example on the input 42. This
execution differs from the one in \cref{table:simple-model-run} in the order
in which the race condition is resolved. Here, the write terminates first, and
only then does the read occur. Note that in this execution there is no
residual execution, and hence no broken promises.} \label{table:simple-model-run-variant}
\endfirsthead
\endhead
\toprule
\# & loc & command & state & unresolved promises & comment\\
\midrule
0 & l\ref{line:main-serverless-function} & main(event) &  & & \\
 &  &  & \texttt{val -> undefined} & & \\
 &  &  & \texttt{hash -> undefined} & & \\
 &  &  & \texttt{event -> \{val: 42\}} & & \\
\midrule
1 & l\ref{line:extract-val} & val = event.val &  & & \\
 &  &  & \texttt{val -> 42} & & \\
 &  &  & \texttt{hash -> undefined} & & \\
 &  &  & \texttt{event -> \{val: 42\}} & & \\
\midrule
2 & l\ref{line:compute-hash} & hash = computeHash(val) &  & & \\
 &  &  & \texttt{val -> 42} & & \\
 &  &  & \texttt{hash -> \(\mathcal{H}(42)\)} & & \\
 &  &  & \texttt{event -> \{val: 42\}} & & \\
\midrule
3 & l\ref{line:connect} & db.connect(\ldots{}) &  & \{$p_{12}$ (db.connect)\} & connection\\
 &  &  &  &  & started\\
\midrule
4 & l\ref{line:connect} & db.connect(\ldots{}) &  &  & connection\\
 &  &  &  &  & established\\
\midrule
5 & l\ref{line:store} & conn.write(\{val, hash\}) &  & \{$p_{13}$ (conn.write)\} & write \\
 &  &  &  &  & started \\
\midrule
6 & l\ref{line:store} & conn.write(\{42, \(\mathcal{H}(42)\)\}) &  &  & write \\
 &  &  &  &  & finished \\
\midrule
7 & l\ref{line:db-second-connect} & db.connect(\ldots{}) &  & \{$p_{16}$ (db.connect)\} & connection\\
 &  &  &  &  & started\\
\midrule
8 & l\ref{line:db-second-connect} & db.connect(\ldots{}) &  &  & connection\\
 &  &  &  &  & established\\
\midrule
9 & l\ref{line:db-read} & conn.read(...) &  & \{$p_{17}$ (con.read)\} & read \\
 &  &  &  &  & started \\
\midrule
10 & l\ref{line:db-read} & conn.read(...) &  &  & read \\
 &  &  &  &  & finished \\
\midrule
11 & l\ref{line:return-stored-n-hash} & () => (\{stored, hash\}) &  & \{$p_{18}$ (produce response)\} & response \\
&  &  &  &  & started \\
\midrule
12 & l\ref{line:return-stored-n-hash} & (\{$\mathcal{S}$,\(\mathcal{H}(42)\)\}) &  &  & response \\
&  &  &  &  & produced \\

\bottomrule
\end{longtable}
\end{center}

In the execution described in \cref{table:simple-model-run-variant} the write
operation starts and concludes (steps \#3--\#6) before the read operation
starts. After the write finishes the read operation runs the same way as
before (steps \#7--\#12). In this case, at the end of the execution there are
no promises still unresolved, and so no residual execution remains.
Consequently, there are no broken promises in this case.

Race conditions can lead to different execution outcomes---some of the
runs are correct and others are faulty---which makes them hard to diagnose and debug.
The need to rely on external conditions (timing, network congestion, etc.) to reproduce
the bugs makes the process of fixing the underlying problem much harder. Bugs
resulting from broken promises in serverless applications suffer from
the same complexity, and might in fact be harder to debug since serverless
applications run in ephemeral cloud environments that are supported by few
debugging tools.

\subsection{Formal semantics of the function reuse model}

In practice, serverless platforms do not create a new execution environment
for every function invocation. Doing so would involve needless compute resources being expanded on repeatedly performing the same tasks (creating, starting, and destroying VMs and containers), and
lead to unnecessarily increased latencies for function calls (known as the
\emph{cold-start problem}~\cite{baldini2017serverless}). Instead, platforms reuse execution environments
across events, only disposing of them in order to reclaim resources, for
example when the load decreases such that the total number of available
execution environments can be reduced.

We now extend the formal semantics presented above to obtain the \emph{function reuse model} of
serverless function execution. In this model, execution environments are
reused, and events can be started in an environment that previously processed
another event. All runs admitted by the single execution model are also legal runs under the function
reuse model. However, the function reuse model admits runs that are
not possible under the single execution model.

In this section we describe the modification made to the formal semantics of
the single execution model in order to produce the formal semantics of the
function reuse model.

\begin{figure}
Start Event:
\[
<receive(v), (M, f, Pr)>\Longrightarrow(M[v/input], newEid(), Pr)
\]\vspace{5pt}

End Event:
\[
\inference{M(response)=v & e\in\mathcal{E}_{id}}{<respond(v), (M,e,Pr)>\Longrightarrow(M,f,Pr)}
\]

Invalidate Env:
\[
<\varepsilon, (M,f,Pr)>\Longrightarrow(M,d,Pr)
\]

Local Transition:
\[
\inference{e\in\mathcal{E}_{id}}{<C, (M, e, Pr)>\Longrightarrow(M', e, Pr)}
\]\vspace{5pt}

Start Asynchronous Task:
\[
<startAsync(p), (M, E, Pr)>\Longrightarrow(M, E, Pr\cup\{p\})
\]\vspace{5pt}

End Asynchronous Task:
\[
\inference{p\in Pr}{<resolve(p), (M, E, Pr)>\Longrightarrow(M, E, Pr\setminus\{p\})}
\]\vspace{5pt}

\Description{Semantics of the function ruse model}
\caption{Extending the small-step operational semantics to the function reuse
model of serverless function execution. The End Event rule was modified from
the one found in \cref{sem:existing-simple} and an Invalidate Env rule was
added.}\label{sem:existing-full}
\end{figure}

Figure~\ref{sem:existing-full} shows the modified semantics. The Start Event,
Local Transition, Start Asynchronous Task, and End Asynchronous Task rules are
carried over without modification from the semantics of the single execution
model (\Cref{sem:existing-simple}) 
The changes made include a modified End Event rule, and a new Invalidate
Env rule.

The only change made to the End Event rule is that now instead of changing the
event component of the state ($E$) to $d$, we change it to $f$. This allows
future start events to occur in the same function.

The new Invalidate Env rule is an $\varepsilon$-transition (i.e., a
nondeterministic step) that changes the state of the function from `ready'
(represented by $f$) to `done'/`terminated' (represented by $d$).

The following lemma shows that under the function reuse model event processing
cannot be considered isolated, as residual execution of a past event may
impact a currently running execution:

\begin{lemma}[Cross-event interference]
During the processing of event $e$, there might be some promise $p'\in Pr$ which was started during the processing of a past event $e'$
\end{lemma}

\subsection{Example run of the function reuse model}\label{sec:function-reuse-model-example-run}
We now demonstrate a run of the serverless function in the running example
(\S\ref{sec:example}) under the function reuse model.

\Cref{table:full-model-run} describes a run of the function on two consecutive
events. The inputs for the first and second events are 42 and 112
respectively. Once the first event processing is done, the second event begins
immediately in the same function runtime as the first event. As a result, the
state from the first execution, including values stored in global variables,
and promises of unresolved asynchronous tasks, are carried over.

\begin{center}
\footnotesize
\begin{longtable}{rlllll}
\caption{An execution of the serverless function in the running example under
the function-reuse model. The example run consists of two consecutive
invocations of the serverless function with the input values 42 and 112. Note
that the local state of the function, including pending asynchronous I/O,
carries over from one  execution to the next.}\label{table:full-model-run}
\endfirsthead
\endhead
\toprule
\# & loc & command & state & unresolved promises & comment\\
\midrule
0 & l\ref{line:main-serverless-function} & main(event) &  & & \\
 &  &  & \texttt{val -> undefined} & & \\
 &  &  & \texttt{hash -> undefined} & & \\
 &  &  & \texttt{event -> \{val: 42\}} & & \\
\midrule
1 & l\ref{line:extract-val} & val = event.val &  & & \\
 &  &  & \texttt{val -> 42} & & \\
 &  &  & \texttt{hash -> undefined} & & \\
 &  &  & \texttt{event -> \{val: 42\}} & & \\
\midrule
2 & l\ref{line:compute-hash} & hash = computeHash(val) &  & & \\
 &  &  & \texttt{val -> 42} & & \\
 &  &  & \texttt{hash -> \(\mathcal{H}(42)\)} & & \\
 &  &  & \texttt{event -> \{val: 42\}} & & \\
\midrule
3 & l\ref{line:connect} & db.connect(\ldots{}) &  & \{$p_{12}$ (db.connect)\} & connection\\
&  &  &  &  & started\\
\midrule
4 & l\ref{line:db-second-connect} & db.connect(\ldots{}) &  & \{$p_{12}$ (db.connect), & connection\\
 &  &  &  & $p_{16}$ (db.connect)\} & started\\
\midrule
5 & l\ref{line:db-second-connect} & db.connect(\ldots{}) &  & \{$p_{12}$ (db.connect)\} & connection\\
 &  &  &  &  & established\\
\midrule
6 & l\ref{line:db-read} & conn.read(...) &  & \{$p_{12}$ (db.connect), & read \\
 &  &  &  & $p_{17}$ (con.read)\} & started \\
\midrule
7 & l\ref{line:db-read} & conn.read(...) &  & \{$p_{12}$ (db.connect)\} & read \\
 &  &  &  &  & finished \\
\midrule
8 & l\ref{line:return-stored-n-hash} & () => (\{stored, hash\}) &  & \{$p_{12}$ (db.connect), & response \\
 &  &  &  & $p_{18}$ (produce response)\} & started \\
\midrule
9 & l\ref{line:return-stored-n-hash} & (\{$\mathcal{S}$, $\mathcal{H}(42)$\})  &  & \{$p_{12}$ (db.connect)\} & response \\
 &  &  &  &  & produced \\
\toprule
10 & l\ref{line:main-serverless-function} & main(event) &  & \{$p_{12}$ (db.connect)\} & promise $p_{12}$ carried\\
 &  &  & \texttt{val -> 42} & & over from\\
 &  &  & \texttt{hash -> \(\mathcal{H}(42)\)} & & previous run \\
 &  &  & \texttt{event -> \{val: 112}\} & & \\
\midrule
11 & l\ref{line:extract-val} & val = event.val &  & \{$p_{12}$ (db.connect)\} & \\
 &  &  & \texttt{val -> 112} & & \\
 &  &  & \texttt{hash -> \(\mathcal{H}(42)\)} & & \\
 &  &  & \texttt{event -> \{val: 112}\} & & \\
\midrule
12 & l\ref{line:connect} & db.connect(\ldots{}) &  &  & connection\\
 &  &  &  &  & established\\
\midrule
13 & l\ref{line:store} & conn.write({112, \(\mathcal{H}(42)\)}) &  & \{$p_{13}$ (con.write)\} & write \\
 &  &  &  &  & started \\
\midrule
14 & l\ref{line:store} & conn.write(...) &  &  & write \\
 &  &  &  &  & finished \\
\midrule
15 & l\ref{line:compute-hash} & hash = computeHash(val) &  & & \\
 &  &  & \texttt{val -> 112} & & \\
 &  &  & \texttt{hash -> \(\mathcal{H}(112)\)} & & \\
 &  &  & \texttt{event -> \{val: 112}\} & & \\
\midrule
16 & l\ref{line:connect} & db.connect(\ldots{}) &  & \{$p_{12}$ (db.connect)\} & connection\\
 &  &  &  &  & started\\
\midrule
17 & l\ref{line:db-second-connect} & db.connect(\ldots{}) &  & \{$p_{12}$ (db.connect), & connection\\
 &  &  &  & $p_{16}$ (db.connect)\} & started\\
\midrule
18 & l\ref{line:db-second-connect} & db.connect(\ldots{}) &  & \{$p_{12}$ (db.connect)\} & connection\\
 &  &  &  &  & established\\
\midrule
19 & l\ref{line:db-read} & conn.read(...) &  & \{$p_{12}$ (db.connect), & read \\
 &  &  &  & $p_{17}$ (con.read)\} & started \\
\midrule
20 & l\ref{line:db-read} & conn.read(...) &  & \{$p_{12}$ (db.connect)\} & read \\
 &  &  &  &  & finished \\
\midrule
21 & l\ref{line:return-stored-n-hash} & () => (\{stored, hash\}) &  & \{$p_{12}$ (db.connect), & response \\
 &  &  &  & $p_{18}$ (produce response)\} & started \\
\midrule
22 & l\ref{line:return-stored-n-hash} & (\{$\mathcal{S}$, $\mathcal{H}(112)$\})  &  & \{$p_{12}$ (db.connect)\} & response \\
 &  &  &  &  & produced \\
\bottomrule
\end{longtable}
\end{center}

The first event execution, steps \#0-\#9 are the same as in single-execution
model (\Cref{table:simple-model-run}). In step \#10 the second event processing
starts, with the input 112. Since the execution environment is reused the
global state at the start of the second invocation is the same as it was when
the first one ended. The values stored in the global variables are the ones
that were written in the first invocation, and the asynchronous I/O operations
that started in the first invocation and had not finished are still not
resolved.

In step \#11 the global variable \texttt{val} is updated to hold the value
112. 

In this example we consider a run in which after the variable \texttt{val} is
updated, the connection request carried over from the first invocation
successfully resolves and the rest of the promise chain is processed. Once the
connection is established, the promise is resolved (step \#12).

Before starting the database write operation, the execution evaluates the
object that needs to be stored \lineRef{store}. The stored value consists of
the input, stored in the \texttt{val} global variable, and the hash value,
stored in  the \texttt{hash} global variable. In step \#13, the value stored
in the global variable \texttt{val} is the up-to-date value 112, but the value
stored in the global variable \texttt{hash} is $\mathcal{H}(42)$, a stale
value carried over from the first invocation. In steps \#13--\#14 the
execution writes the object \texttt{\{112, $\mathcal{H}(42)$\}} to the database.
The value stored in the database is now incorrect.

The rest of the example execution, namely steps \#15--\#22, are similar to
the single-execution model (\Cref{table:simple-model-run}).

This example shows three problems that occur in the function reuse model. (i)
There is residual execution at the end of the second run; if a third run does
not start, this will result in a broken promise. (ii) During the second
function execution, we saw processing of code that started in the first
execution, namely steps \#12--\#14; violating isolation. (iii) The data
written to the database was inconsistent, partly reading stale data, and
partly reading updated data, violating correctness.

\subsection{Additional considerations}

\paragraph{Timeout-sensitive APIs}

Since serverless platform do not perform processing when idle, the execution
environment is stopped once the function terminates. Any residual execution
will only run when the next event arrives and the execution environment is
allowed to run again. This creates periods of time in which, to external
observers, the process appears to be dead.

Consider for example the \texttt{db.connect} operation in the example. This
call starts a connection with the database service. In order to establish a
connection, the process and the database communicate via some protocol.
However, if in the middle of this communication the process stops responding
to messages from the server, after a certain period of time the server might
conclude that the process it is communicating with is no longer available, and
terminate the protocol without establishing a connection.

Once the process resumes execution, the asynchronous task will result in an
error.

\paragraph{Other multi-threading models}
In this work we use the terminology of promises and asynchronous operations,
out of convenience and for clarity of presentation. The same model and the
same rules apply to regular multi-threading, with the Start Asynchronous Task
and End Asynchronous Task rules representing the creation and elimination of a
program thread.

\paragraph{Security considerations}

Finally, while it is outside of the scope of this work, it is worth noting
that when residual execution is carried over across invocation, this may pose
privacy and security problems. For example, if the hash value computed in the
code in \cref{code:main-handler} depends on secret information, then the
residual execution of the previous event might leak that information by
writing it to the database with the wrong access permissions, allowing the
originator of a request to access private data they should not be able to
view.

Indeed, works focusing on information security in serverless applications
(\cite{alpernas2018secure}) restrict function reuse to ensure that no state or
residual execution are carried over across function invocations.

\section{The \emph{Correct} Semantics of Serverless Function Termination}\label{sec:prop-semantics}

In this section, we present two approaches to addressing the problems presented in \cref{sec:curr-semantics}.
The first approach (\S\ref{sec:strawman}) does not require changes to serverless platforms, but instead changes the correctness requirements
imposed on applications. However, this approach adds to the response latency
of serverless applications and thus reduces their utility. We address this in our second approach 
(\S\ref{subsec:terminalsep}), which requires changing the serverless platform---to introduce
a new mechanism that applications use to signal that event processing is complete---and 
applications, but addresses the problem without adding to response latency. 

\subsection{Approach 1: Waiting for all promises to resolve}\label{sec:strawman}

The problems presented in \cref{sec:curr-semantics} are caused by the
existence of residual execution once the serverless function terminates. This
residual execution may end up a broken promise, or may interfere with future
executions of the serverless function, depending on things such as the load on
the serverless system.

A straightforward approach to ensuring that no residual execution exists once
the serverless function terminates is requiring that termination steps may
only occur when no unresolved promises exist. \Cref{sem:straw-man-semantics}
describes the modification made to the formal semantics of the function reuse
model in order to guarantee freedom from residual execution.

\begin{figure}
End Event:
\[
\inference{M(response)=v & e\in\mathcal{E}_{id}}{<respond(v), (M, e, \emptyset)>\Longrightarrow(M,f,\emptyset)}
\]\vspace{5pt}

\Description{Semantics for the naive solution}
\caption{Modified formal semantics for serverless function executions that
only terminate once all asynchronous operations have been resolved.}\label{sem:straw-man-semantics}
\end{figure}

The only change made to the semantics in \cref{sem:straw-man-semantics} is to
the End Event rule. In the modified semantics, the premise of the End Event
rule requires that the set of unresolved promises $Pr$ be empty. This
guarantees that termination only occurs once all promises have been resolved.

In the case of the example in \cref{code:main-handler}, this means that
termination will now always wait for the entire promise chain of the write
operation (\crefrange{line:connect}{line:catch-err}) before terminating the
function. For example, the run described in \cref{table:simple-model-run} is
not legal under these semantics, while the run described in
\cref{table:simple-model-run-variant} is still legal.

Under this solution, sending a response to the caller will always depend on
the resolution of all asynchronous operations started during the execution.
However, in many cases the choice to perform some operation in the background
and not wait for the operation to finish before sending a response is
\emph{intentional}. When working in an interactive application, the response
latency has significant impact on user experience \cite{arapakis2014impact,tolia2006quantifying,attig2017system}, and
sending the response back to the user as soon as possible has a high priority.
For that reason, developers often choose to perform parts of the computation
(including bookkeeping and other operation not directly related to the
response data) after the response has been sent, so as to not delay the
response more than is strictly necessary.

In the example in \cref{code:main-handler}, the developer intentionally choses
to write the value and hash to the database
(\crefrange{line:connect}{line:store}) without waiting for the operation to
resolve before sending the response back to the
user (\cref{line:return-stored-n-hash}).

\subsection{Approach 2: Separating response from termination}\label{subsec:terminalsep}

The approach described in the previous subsection adds to the serverless
application's response latency. Here we describe an alternate approach that
avoids this additional latency by decoupling response events and termination
events. Response events send a message back to the originator of
the event, while termination events signal that processing is complete.

The addition of an explicit termination action requires a change in semantics
but also a change in syntax. We add a new kind of command to the program,
\texttt{end()} and require that developers explicitly use the new command to
denote function termination. 

\begin{figure}
Start Event:
\[ 
<receive(v), (M, f, \emptyset)>\Longrightarrow(M[v/input], newEid(), \emptyset)
\]\vspace{5pt}

Respond:
\[
\inference{M(response)=v & e\in\mathcal{E}_{id}}{<respond(v), (M, e, Pr)>\Longrightarrow(M, e, Pr)}
\]\vspace{5pt}

End Event:
\[
\inference{e\in\mathcal{E}_{id}}{<end(), (M,e,\emptyset)>\Longrightarrow(M,f,\emptyset)}
\]\vspace{5pt}

Invalidate Env:
\[
<\varepsilon, (M,f,Pr)>\Longrightarrow(M,d,Pr)
\]\vspace{5pt}

Local Transition:
\[
\inference{e\in\mathcal{E}_{id}}{<C, (M, e, Pr)>\Longrightarrow(M', e, Pr)}
\]\vspace{5pt}

Start Asynchronous Task:
\[
<startAsync(p), (M, E, Pr)>\Longrightarrow(M, E, Pr\cup\{p\})
\]\vspace{5pt}

End Asynchronous Task:
\[
\inference{p\in Pr}{<resolve(p), (M, E, Pr)>\Longrightarrow(M, E, Pr\setminus\{p\})}
\]
\Description{Semantics for the correct model}
\caption{The proposed semantics for correct serverless function execution.}\label{sem:proposed}
\end{figure}

Figure~\ref{sem:proposed} describes our proposal for a small-step operational
semantics of the full model of serverless function execution. This semantics
modifies the End Event rule, and adds a new Respond rule.

The changed End Event rule now requires that the set of promises $pr$ be empty
when ending the function run. Additionally, the End Event no longer sends a
response to the requestor.

The new Respond event sends a message with the value $v$ to the originator of
the event.

The following theorems prove that the proposed semantics fix the broken promises and interference problems:

\begin{theorem}[No broken promises]
Every promise started during the execution of a serverless function is either guaranteed to be resolved, or the function will not end
\end{theorem}

\begin{theorem}[No cross-event interference]
Every promise $p\in Pr$ running during the processing of some event $e$ was also created during the processing of event $e$
\end{theorem}

\subsection{Modified program}\label{sub:correct-example-run}

\Cref{code:correct-main-handler} shows a variant of the example in
\cref{code:main-handler}, modified to run under the semantics defined in
\cref{sem:proposed}.

\begin{figure}
\footnotesize
\begin{lstlisting}[style=js]
const cloudProvider = require('...') (*@ \label{line:correct-cloud-provider-require} @*)
const db = cloudProvider.DB()        (*@ \label{line:correct-cloud-provider-db} @*)

let val, hash                        (*@ \label{line:correct-global-vars} @*)

function computeHash(val) {...}      (*@ \label{line:correct-hash-func-impl} @*)

function main (event) => {     (*@ \label{line:correct-main-serverless-function} @*)
  val = event.val                    (*@ \label{line:correct-extract-val} @*)
  hash = computeHash(val)            (*@ \label{line:correct-compute-hash} @*)

  pr_write = db.connect(...) // returns a promise         (*@ \label{line:correct-connect} @*)
    .then((con) => con.write({val, hash})      (*@ \label{line:correct-store} @*)
    .catch((err) => ...)                       (*@ \label{line:correct-catch-err} @*)
    
  pr_read = db.connect(...)                       (*@ \label{line:correct-db-second-connect} @*)
    .then((con) => con.read(...))         (*@ \label{line:correct-db-read} @*)
    .then((stored) => ({stored, hash}))   (*@ \label{line:correct-return-stored-n-hash} @*)
  
  Promise.allSettled([pr_write, pr_read])    (*@ \label{line:correct-promise-all} @*)
  	.finally(() => end())             (*@ \label{line:correct-terminate} @*)

  return pr_read                      (*@ \label{line:correct-respond} @*)
}
\end{lstlisting}
\Description{Code for the correctly modified running example}
\caption{A variant of the serverless function from the example in
\cref{code:main-handler} modified to work under the correct semantics. In this
variant, the response, sent to the user via the \texttt{return} statement, is
kept the same as in the original. However, the two promise chains created by
the execution are combined (\cref{line:correct-promise-all}), and after they
are resolved the \texttt{end()} command is called (\cref{line:correct-terminate}).}\label{code:correct-main-handler}
\end{figure}

The modified program now stores the promises created by the write and read
asynchronous operations in local variables
(\cref{line:correct-connect,line:correct-db-second-connect}). The program then
combines these two promises using a \texttt{Promise.allSettled()} combinator
(\cref{line:correct-promise-all}). The \texttt{Promise.allSettled()}
combinator returns a promise that resolves when all the constituent promises
resolve (either fulfilled or rejected). In \cref{line:correct-terminate} a
\texttt{.finally()} call ensures that after the combined promise resolves,
regardless of whether there is an exception or it resolves correctly with a
value, the \texttt{end()} call is made, representing the end of all
asynchronous operations started by this function execution. The call to
\texttt{end()} represents the termination of the program. While the
asynchronous operations are executing, the function responds to the caller
with the correct value---the resolution of the read asynchronous
operation---via the returned value of the program
(\cref{line:correct-respond}).

\begin{figure}
    \centering
    \includegraphics[width=0.95\textwidth]{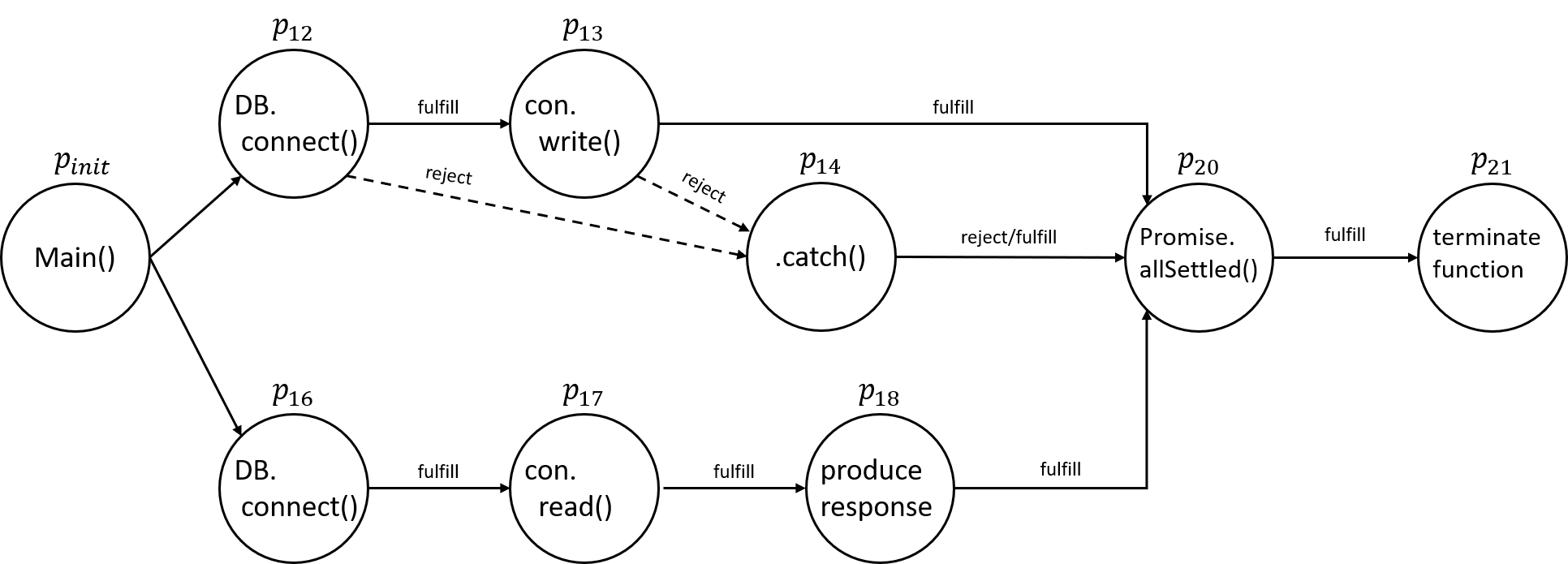}
    \Description{Promise graph for the correctly modified running example}
    \caption{The promise graph of the main function variant described in \cref{code:correct-main-handler}. Note that $p_{18}$ now only produces the response, and $p_{21}$ is now the explicit \texttt{end} call.}
    \label{fig:correct-example-promise-graph}
\end{figure}

\Cref{fig:correct-example-promise-graph} shows the promise graph of the
modified serverless function from \cref{code:correct-main-handler}. Whereas
previously we had a single node representing both response and termination
($p_{18}$), in this promise graph, we now have two separate nodes for
response and for termination, $p_{18}$ and $p_{21}$ respectively.

We can see that in the current promise graph both chains of asynchronous
operations lead, via the \texttt{Promise.allSettled()} combinator (node $p_{20}$) to
termination node $p_{21}$, guaranteeing that all asynchronous operations are
resolved before the function terminates. There is still a race condition
between the database write operation ($p_{13}$) and the production of the response sent to the
caller ($p_{18}$). However, since the response to the
user is no longer the marker for termination, this race condition is no longer
a problem, indeed it is the intended program behaviour.

\section{Implementation}\label{sec:implementation}

We have implemented our proposed semantics in a modified version of the
OpenWhisk(\cite{openwhisk-github-repo}) serverless platform for serverless
functions running on the Node.js runtime.

In OpenWhisk, function runtimes are implemented in docker containers, and are
orchestrated and managed by a component called OpenWhisk Invoker. The Invoker
manages the lifecycle of the container instances, creating and destroying
containers as necessary; it communicates with the instances via an http API.
The http API consists of two messages---\texttt{init} and \texttt{run}. 

\paragraph{Modifications to the OpenWhisk Invoker}
Upon the arrival of an event, the Invoker makes an \texttt{init} call to one
of the available container instances (it creates a new one if none is
available). The \texttt{init} call is \emph{blocking}, and the Invoker waits
for a response from the container to notify it that whatever initialization
code needs to run has finished running. The \texttt{init} call is made
regardless of whether the container image is new (cold-start) or existing
(warm-start).

Once the \texttt{init} call is done, the Invoker makes a \texttt{run} call,
passing the event as a parameter of the call. The \texttt{run} call is also a
\emph{blocking} call. The \texttt{run} call starts the function execution in
the container, and once a response is produced (i.e., the serverless function
returns a value) the response is sent as a response to the \texttt{run} call
and is received by the Invoker.

We implemented our proposed semantics by adding an additional, third, http
API---\texttt{await}. The \texttt{await} call is also a \emph{blocking} call.
Once a response returns from the \texttt{run} call, that response is forwarded
back to the caller of the serverless function. After the response is sent
back, we make an \texttt{await} call. The \texttt{await} call returns once an
\texttt{end} statement was processed in the function run, marking the end of
all asynchronous processing started by the function.

After the \texttt{await} call ends the Invoker proceeds to perform the regular
bookkeeping performed by OpenWhisk (e.g., collecting logs), and either stop
the container image, or process the next event if one is available.

\paragraph{Modifications to the function runtime}

In order to support the additional http API, the language runtime needs to
implement an \texttt{await} http endpoint. We have modified the container
images of the Node.js runtimes currently supported by OpenWhisk---Node.js
versions 10, 12, and 14.

We added an \texttt{end()} function to the JavaScript \texttt{global} object.
Calling the \texttt{end()} function resolves a promise object that is unique
to the current execution of the function. The \texttt{await} http call waits
for the promise object to resolve before sending a response back to the
Invoker. If the promise object is resolved before the \texttt{await} call is
made, for example in case the \texttt{end()} function is called during the
function execution when all asynchronous operation end, then the wait for the
promise to resolve is instantaneous, and the http call returns immediately.
This mechanism takes advantage of the fact that a promise is resolved once,
and any future accesses to the resolved data occur immediately.

\paragraph{Evaluation}

We have measured the runtime overhead of running the modified semantics on
functions that did not have residual execution in the original semantics. We
measured the time between the \texttt{await} http call start and the time a
response was received by the OpenWhisk invoker. The average time of the
\texttt{await} http call in cold-start function executions was 25ms, and the
average time in warm-start function executions was 8ms.

We do not measure overheads for cases where there was residual execution in
the original semantics, since in these cases the execution itself changes
rendering the comparison moot.

All measurements were performed on a Lenovo X1-Carbon (6th Generation)
machine, with an Intel® Core™ i7-8550U CPU, 16GB of RAM and an NVMe SSD,
running Ubuntu 20.04.2 LTS.

\subsection{Design considerations}\label{sec:design}

We now discuss several design consideration that needed to be addressed during
the implementation of the proposed semantics.

\paragraph{Timeouts and billing}

Serverless platform bill for processing time only, taking care not to bill for
idle time between events. Additionally, every function execution has a
mandatory time limit for processing, after which it is terminated. Since the
asynchronous processing performed by a function execution is essentially part
of the execution, the time after the \texttt{run} call ends and until the
\texttt{await} call ends, essentially the residual execution of the function,
needs to count both towards billing and towards the function timeout.

\paragraph{Error handling}

In the current semantics, because the response to the user marks the end of
the function execution, there is a guarantee that if any errors occur during
the execution, they occur before the function termination. As a result, if an
error occurs during the execution, it can be sent back to the caller, and the
caller can implement error handling. However, this also poses problems as in
some cases the error sent back to the caller may be caused by asynchronous
execution started by another caller in another event (this may also pose a
security risk).

In the proposed semantics, we may have cases where a response was sent to the
caller, indicating a seemingly successful execution, only to be followed by an
error occurring during the \texttt{await} phase of the function execution.
These errors are not reported back to the caller. Instead, it becomes critical
that developers implement error handling mechanisms at the serverless function
level.

\paragraph{JavaScript vs. other language runtimes}

In our implementation we chose to focus on JavaScript runtimes. Asynchronous
execution is the main tool in JavaScript for performing any kind of I/O, and
the asynchronous model and promises in general pose a challenges to novice and
experienced developers alike. In conjunction with the popularity of Node.js as
a serverless function runtime, we expect the problems presented in this paper
to be most prevalent in serverless function written in JavaScript.

However, the problems presented here are by no means unique to JavaScript.
Most popular runtimes have asynchronous I/O facilities, and support for
promises or promise-like mechanisms (e.g., futures). Furthermore, the problems
presented here are not limited to asynchronous I/O, and can occur even when
working with `vanilla' threads.

Adding support for other language runtimes will require implementing the
\texttt{await} http API in the relevant container image. This will typically
require adding a way to make a global function call from the executing code,
and a marking mechanism for setting the processing status of the function to
finished. The details will depend on the specifics of the language runtime.

\paragraph{Other serverless platforms} 

Since all other major serverless platform, apart from OpenWhisk, are
proprietary, it is impossible for us to determine the exact mechanism by which
function executions are managed. Of course, making any changes to the
platforms is also outside the scope of our reach. However, based on the
observed behaviour of these platforms, as well as published documentation, for
example AWS instructions for defining custom language runtimes \cite{aws-custom-lambda}, we
strongly suspect that they employ similar mechanisms to the ones in OpenWhisk,
and that similar modifications can be made to other platforms to support our
proposed semantics.

Unfortunately, due to the inherent coupling of response and termination in the
existing semantics of serverless platforms, it is impossible to design a
language- or library-based implementation for the semantics proposed in
Approach 2 (\cref{subsec:terminalsep}). However, the semantics proposed in
Approach 1 (\cref{sec:strawman}) can be implemented using a strictly language
based approach, without requiring modifications to the serverless platform,
making them applicable for deployment in proprietary serverless platforms.

\section{Related Work}

\paragraph{Serverless semantics}

Several formal models have been proposed for describing serverless computing
\cite{jangda2019formal,obetz2020formalizing,gabbrielli2019no}, addressing the
communication mechanisms, load balancing, cold- and warm-starts of serverless
functions, error handling, function composition, direct function invocation,
and other aspects of serverless computing platforms. However, none of these
works explicitly address the termination condition of function and the local
state of function execution. Indeed, the properties described in our works are
not expressible in the semantics presented in previous work. In this work we
explicitly focus on the local state and termination conditions of serverless
functions. Our model abstracts away many of the details present in previous
work, such as load-balancing and handling of function execution errors.

\paragraph{Analysis of serverless functions}

Past work on analysis of serverless functions identified multiple potential
correctness and security problems presented by serverless platforms.
\cite{winzinger2019model} propose a model based approach to analyzing
behaviour of concurrent serverless function execution. \cite{obetz2019static}
construct call graphs to reason about serverless applications.
\cite{alpernas2018secure} and \cite{datta2020valve} propose a language-based
and serverless platform-based runtime system for enforcing information flow
control in serverless applications. To the best of our knowledge, ours is the
first work to detect and provide a solution for the termination problem in
serverless applications.

\paragraph{Analysis of asynchronous executions}

Both static and dynamic analysis tools have been proposed for reasoning about
and detecting problems with asynchronous executions in JavaScript programs.
\cite{madsen2017model} proposes a formalism for describing the causal
relations of promises in JavaScript programs and describes several classes of
correctness violations that can be detected by analyzing the promise graph of
a program. \cite{alimadadi2018finding} extends this work with an runtime
analysis tool that produces promise graphs for programs. We see these works as
complementary to this work, as similar approaches can be used to automate the
placement of the \texttt{end} calls in serverless functions.

\section{Conclusion}\label{sec:conclusion}

The way serverless platforms determine when a function terminates has deep
impacts on application execution. When functions perform asynchronous I/O, as
is almost always the case in JavaScript functions, the existing semantics that
tie termination to function response lead to inconsistent runs, developer
confusion, and corrupted data. The na\"ive approach to fixing these problems
involves delaying the response until after all asynchronous operations are
done, and leads to significantly increased function latency.

In this work we showed a new approach to function termination, that fixes the
issues caused by the existing way platforms decide on termination, without
impacting function latency. Our proposal separates function response from
function termination, allowing functions to continue execution after sending a
response, and terminate the execution explicitly when all asynchronous
processing is finished. In addition to fixing existing problems, our model
gives developers the ability to intentionally and correctly perform processing
`in the background' allowing for application with improved perceived latency
and usability in interactive applications. 

Finally, we showed an implementation of our proposed semantics in a modified
version of the OpenWhisk serverless platform and Node.js v10,v12, and v14
runtimes. Our implementation adds an overhead of just 8ms to serverless
function runs that do not have residual execution (25ms for cold-start
executions).

\bibliography{refs}

\end{document}